\documentclass[prb,preprint]{revtex4-1} 

\usepackage{amsmath}  % needed for \tfrac, \bmatrix, etc.
\usepackage{amsfonts} % needed for bold Greek, Fraktur, and blackboard bold
\usepackage{graphicx} % needed for figures
\usepackage{changes}
\newcommand{\ket}[1]{\mbox{$|{ #1}\rangle$}}
\newcommand{\bra}[1]{\mbox{$\langle { #1}|$}}

\begin{document}

\title{An Easier-To-Align Hong-Ou-Mandel Interference Demonstration}
\author{Nicholas S. DiBrita}
\altaffiliation[Present Address: ]{Department of Physics and Astronomy, Rice University, Houston, TX 77251, USA.}
\author{Enrique J. Galvez}
\email{egalvez@colgate.edu} % optional
\affiliation{Department of Physics, Colgate University, Hamilton, NY 13346, USA}

\date{\today}

\begin{abstract}
The Hong-Ou-Mandel interference experiment is a fundamental demonstration of nonclassical interference and a basis for many investigations of quantum information. This experiment involves the interference of two photons reaching a symmetric beamsplitter. When the photons are made indistinguishable in all possible ways, an interference of quantum amplitudes results in both photons always leaving the same beamsplitter output port. Thus, a scan of distinguishable parameters, such as the arrival time difference of the photons reaching the beamsplitter, produces a dip in the coincidences measured at the outputs of the beamsplitter. The main challenge for its implementation as an undergraduate laboratory is the alignment of the photon paths at the beamsplitter. We overcome this difficulty by using a pre-aligned commercial fiber-coupled beamsplitter. In addition, we use waveplates to vary the distinguishability of the photons by their state of polarization. We present a theoretical 
description at the introductory quantum mechanics level of the two types of experiments, plus a discussion of the apparatus alignment and list of parts needed.
\end{abstract}

\maketitle % title page is now complete

\section{Introduction}\label{sec:intro} % Section titles are automatically converted to all-caps.

In 1987 C.K. Hong, Z.Y. Ou, and L. Mandel reported on one of the most consequential experiments in quantum optics.\cite{HOM} It is an experiment that demonstrates the ensuing quantum interference of two properly prepared photons after each arrives separately at an adjacent input 
 port of a symmetric beamsplitter. When all of the properties of the two photons are identical, a striking phenomenon appears: the two photons always exit together at the same output 
 port of the beamsplitter and never exit at separate output 
  ports. This effect is a purely nonclassical phenomenon. The proper way to understand it is from a quantum-mechanical perspective, where the amplitudes for the various possibilities interfere. This result mimics a form of interaction between photons, but one that is solely due to quantum effects, similar to the exchange interaction of electrons in atoms. This quantum interaction has been used for a number of purposes,\cite{Bouchard20} such as entanglement,\cite{Ou88,Shih88} entanglement swapping,\cite{Pan98} teleportation,\cite{Bouwmeester97} implementation of CNOT gates,\cite{OBrien03} and ultimately, quantum computing with photons.\cite{kokRMP07}

\begin{figure}[h!]
\centering
\includegraphics[width=4in]{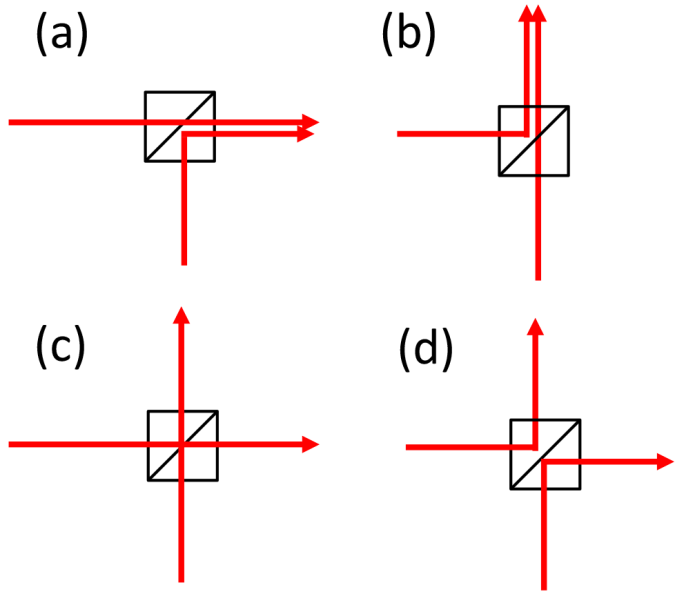}
\caption{Schematic of the four possible paths of two photons, each incident separately on adjacent 
input ports of a beamsplitter.}
\label{fig:HOM}
\end{figure}

The essence of the Hong-Ou-Mandel (HOM) interference phenomenon is shown in Fig.~\ref{fig:HOM}. When two photons arrive separately at adjacent input 
ports of a beamsplitter, there are four possible outcomes. Either the two photons exit together out of the same output 
port in one of two possible ways, as shown in Figs.~\ref{fig:HOM}(a) and \ref{fig:HOM}(b), or they exit out of separate 
ports in one of two possible ways, as shown in Figs.~\ref{fig:HOM}(c) and \ref{fig:HOM}(d). Following Feynman,\cite{Feynman} consider the event when both photons exit out of separate output 
ports of the beamsplitter. If the photons are indistinguishable, the probability for the event is the square of the sum of the probability amplitudes for each possibility considered separately. If the possibilities are distinguishable, then the probability of the event is the sum of the probabilities of the possibilities. 

Now assume the beamsplitter to be a {\em symmetric} one, i.e., with equal probabilities to transmit and reflect light, and equal amplitudes for reflection and transmission from either side of the beamsplitter. It is common to call the probability amplitudes for transmission and reflection $t$ and $r$, respectively. The absolute value for both $t$ and $r$ has to be $1/\sqrt{2}$, so that the probability of transmission and reflection is $1/2$ in each case. However, to conserve energy, or equivalently, probability, the transmission and reflection amplitudes have to be out of phase by $\pi/2$ for the case of the symmetric beamsplitter.\cite{Zeilinger,Holbrow} It is common to attach this phase to the reflection amplitude, so $r=\exp(i\pi/2)/\sqrt{2}=i/\sqrt{2}$ and $t=1/\sqrt{2}$. The probability amplitude that both photons come out of separate %sides
output ports of the beamsplitter has two terms: when both transmit, it is $tt=1/2$  [Fig.~\ref{fig:HOM}(c)]; and both reflect, it is $rr=-1/2$  [Fig.~\ref{fig:HOM}(d)]. The probability for the event is then
\begin{equation}
P_{\rm ind}=|tt+rr|^2=0.
\label{eq:homint}
\end{equation}
That is, the two possibilities interfere destructively. 

If the photons are distinguishable, such as when they arrive at the beamsplitter at distinguishable different times, then the probability is
\begin{equation}
P_{\rm dis}=|tt|^2+|rr|^2=1/2.
\end{equation}
Distinguishable different times means that a measurement of the two arrival times of the photons can be used to distinguish between the two possibilities. Other distinguishing attributes are the photons' polarization, energy, or spatial mode.

We note that the previous analysis applies to bosons, like the photon. For fermions (for example, electrons), the amplitude rule of Eq.~(\ref{eq:homint}) is not a sum but a difference of the two probability amplitudes.\cite{Feynman} This fact is due to the exchange symmetry of indistinguishable fermions, which unlike bosons, cannot occupy the same state (i.e., both fermions having the same momentum). Thus, in the HOM experiments with electrons,\cite{BocquillonSci13} the probability of Eq.~(\ref{eq:homint}) is 1. Feynman explains the distinction between bosons and fermions with a similar type of experiment, of identical particles in a head-on collision.\cite{Feynman} This phenomenon is more formally described in terms of the symmetry of the two-particle wavefunction, presented in Sec.~\ref{sec:results}. Ultimately, the HOM experiment is a demonstration of the superposition of the state of two particles and how it leads to measurable interference effects that are purely quantum
 mechanical.
 
Recreation of this demonstration is not straightforward, mostly because the experimental alignment requires much effort and expertise, and thus is time consuming. To see the interference, both photons created from the same source---spontaneous parametric down-conversion (described below)--- have to travel exactly the same distance to the beamsplitter, so setting up the photon paths needs very careful alignment. Additionally, the experiment requires hardware that facilitates scanning the photon path difference by tens of micrometers. 
A final challenge occurs at the beamsplitter. The photons' spatial mode must fully overlap at the beam splitter and along the output paths. Otherwise they will be spatially distinguishable. For educational purposes, this demonstration has been done before in free space,\cite{GarivotoEJP12} where the experimentalists implemented the following clever method of alignment: a Mach-Zehnder interferometer was set up such that its separate arms were the two photon paths. Once the paths were aligned to be equal via interferometry, the first beamsplitter was replaced by the down-conversion crystal. 

The purpose of this article is to report on a Hong-Ou-Mandel interference demonstration that eliminates the free-space alignment of the photon beams reaching the beamsplitter. Instead, we use a commercial pre-aligned device.
The cost of this device is within the norm for component hardware in common use in quantum optics instructional physics laboratories. 
This new system adds its own complication, though. That is, the down-converted photons need to be coupled to single-mode fibers. We find this extra effort to be a worthwhile trade-off. In Appendix A, we provide a suggested procedure for the required alignment. We also note that this experiment is already sold commercially as a black-box-type experiment.\cite{Qubitekk,Qutools} Our experiment is based on the beam-splitting component in one of these commercial products, but our aim is a demonstration in which students set up the entire apparatus, and it is done at a lesser cost.

The article is organized as follows. In Sec.~\ref{sec:app}, we provide a detailed description of the experimental apparatus. Section~\ref{sec:results} follows with a quantum-mechanical description of the phenomenon that fits within the curricular formalism of an undergraduate course, along with the experimental results. Two appendices give alignment procedures, a parts list and component costs.

\section{Apparatus}\label{sec:app}

\begin{figure}[h!]
  \centering
\includegraphics[width=5.5in]{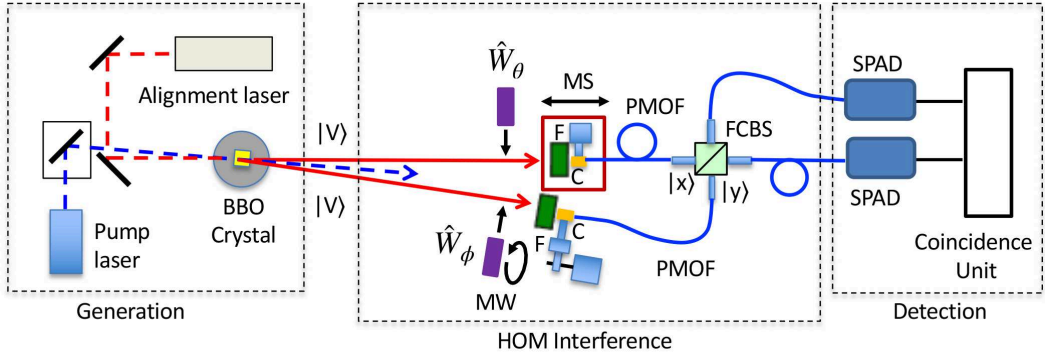}
\caption{Sections of the apparatus. A 405-nm pump laser beam was steered to a BBO crystal to generate vertically polarized photons. A fiber-coupled beamsplitter (FCBS)  provided the photon interference. Other hardware components included half-wave plates ($\hat{W}_\theta$ and $\hat{W}_\phi$) oriented by angles $\theta$ and $\phi$
 to the vertical, respectively, with one on a motorized rotation mount (MW); bandpass filters (F); adjustable-focus collimators (C), with one of them mounted on a motorized translation stage (MS) on top of a manual stage; polarization-maintaining optical fibers (PMOF); and single-photon avalanche diode detectors (SPAD).}
\label{fig:app}
\end{figure}

A diagram of the apparatus is shown in Fig.~\ref{fig:app}. The figure has been sectioned to highlight important parts. The first section is the source of photon pairs via type-I spontaneous parametric down-conversion. A pump laser emitted horizontally polarized light of wavelength 405.4 nm. It was steered toward a beta barium borate (BBO)  crystal that produced photon pairs that were vertically polarized. Up to this point, this is a standard setup for undergraduate quantum optics experiments.\cite{GalPRL14,URL}
In the central section, collimators (C) collect photons into optical fibers. They were placed at $\pm3^\circ$ from the incident pump-beam direction, and located 1-m away from the crystal. Bandpass filters (F), set to collect photons that are near the degenerate wavelength of 810.8~nm, were attached to the collimators along with a mounted iris.
Before we continue with more detail of the central section, we add that the third section is also standard: single-photon avalanche diode detectors (SPAD) fed digital electronic pulses from photon detections to an electronic counting and coincidence unit, which in turn fed data to a laptop/desktop with data acquisition programs written in MATLAB.\cite{URL}

\begin{figure}[h!]
 \centering
\includegraphics[width=3in]{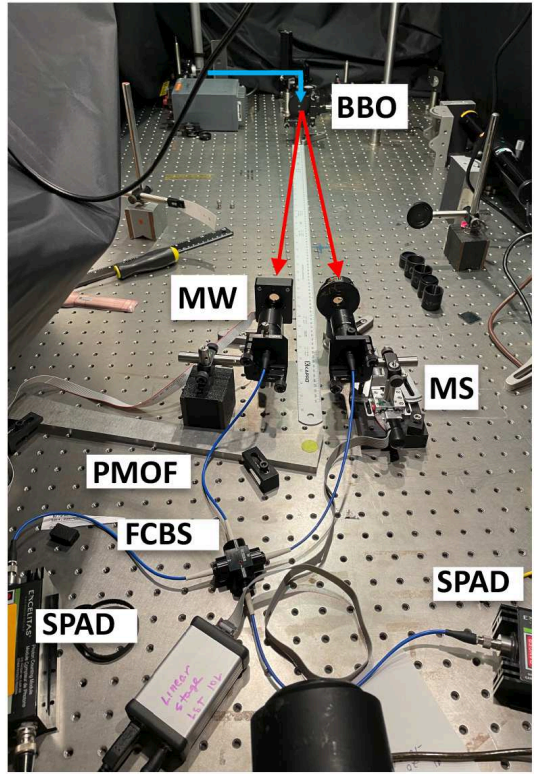}
\caption{Photograph of the apparatus showing the main components: down-conversion crystal (BBO), motorized rotation mount (MW); and linear stage  (MS), polarization-maintaining optical fiber (PMOF); fiber-coupled beamsplitter (FCBS); and photon detectors (SPAD).}
\label{fig:appfig}
\end{figure}

The central portion of the apparatus has several new components. Figure~\ref{fig:appfig} shows a photograph that emphasizes this section of the apparatus. The main component is a commercial pre-aligned fiber-coupled beamsplitter (FCBS). Because the polarization of the light has to be maintained, the fibers are single mode and polarization maintaining (PMOF). The two fiber inputs of the FCBS are connected to the collimators and the fiber outputs are connected to the two SPADs. The first departure from the standard experiments was the use of PMOFs. They significantly restricted the input light. To maximize the coupling of the down-converted photons to the fiber, we used collimators with adjustable focus (C). These collimators have  fiber connectors type FC, which lock the fibers into a specific orientation. To collect photon pairs with the same polarization, both collimators were mounted to have the same orientation of the FC connectors in their mount.

The collimators were mounted on mirror mounts via adapters. One of the collimators' mount was attached to a magnetic mount and placed in contact with an aluminum plate with a 1-m radius of curvature (see Fig.~\ref{fig:app}), with center of curvature located approximately at the position of the crystal. The purpose of the latter was to allow the flexibility to translate the collimator sideways to optimize coincidence counts. The curved path reduced/eliminated the walk-off error that would be introduced if  the collimator was translated linearly sideways. The other collimator's mirror mount was mounted on a double stack of translation stages set to move the collimator toward or away from the crystal. The bottom stage, attached to the breadboard, was a standard manual translation stage with a micrometer screw, whose purpose was to make coarse adjustments. On top of it was a small motorized translation stage (MS) for doing an automatic scan of the crystal-collimator distance. The
 two stages  changed the photon-path difference.

The last component of the arrangement was a pair of half-wave plates. One was mounted on a manual rotation mount and the other  on a motorized rotation mount (MW). They are needed for two purposes: (1) to ensure that the photons enter the optical fibers with the same polarization orientation, and (2) for scanning the polarization distinguishability of the photons as described in Sec.~\ref{sec:pol}.

\section{Two Situations}\label{sec:results}
We describe the experiments in three parts. First, we describe the HOM interference itself. Next, we describe two situations where we can turn the interference on and off, one based on the path length and the other based on the polarization.

\subsection{The HOM Interference}
This HOM interference phenomenon has been described before  in terms of number states.\cite{Kwiat92,GarivotoEJP12} Here, we use an alternative approach in terms of momentum states.
Photons 1 and 2 leaving their place of birth can arrive at the beamsplitter in a state with either momentum $\ket{x}$ or $\ket{y}$, as labeled in Fig.~\ref{fig:app}. Each photon can be in these two possible states. Thus, the full state of both photons is the tensor product of the two photon spaces. Due to the bosonic nature of the quantum state of the two photons, they must be described by a wavefunction that is symmetric by the interchange of the two particles. Thus, the initial state of the two photons is given by
\begin{equation}
\ket{\psi}_i=\frac{1}{\sqrt{2}}\left(\ket{x}_1\ket{y}_2+\ket{y}_1\ket{x}_2\right)
\label{eq:ind}
\end{equation}

To manipulate the state of the light, we can use the matrix notation of quantum mechanics: $\ket{x}=(1\;0)^T$ and $\ket{y}=(0\;1)^T$, with $T$ denoting the transpose of the matrix. The two product states are then given by
\begin{equation}
\ket{x}_1\ket{y}_2= \begin{pmatrix}1 \\ 0\end{pmatrix}_1\otimes \begin{pmatrix}0 \\ 1\end{pmatrix}_2=\begin{pmatrix}0 \\ 1 \\ 0 \\ 0\end{pmatrix}
\end{equation}
and
\begin{equation}
\ket{y}_1\ket{x}_2=\begin{pmatrix}0 \\ 1\end{pmatrix}_1\otimes \begin{pmatrix}1 \\ 0\end{pmatrix}_2=\begin{pmatrix}0 \\ 0 \\ 1 \\ 0\end{pmatrix},
\end{equation}
which results in the initial state given by
 \begin{equation}
\ket{\psi}_i=\frac{1}{\sqrt{2}} \begin{pmatrix}0 \\ 1 \\ 1 \\ 0\end{pmatrix}.
 \end{equation}
 
 The symmetric beamsplitter must apply to each photon space, resulting in an operator acting on the larger space
 \begin{equation}
\hat{B}_2=\begin{pmatrix}t & r \\ r & t\end{pmatrix}\otimes\begin{pmatrix}t & r \\r & t\end{pmatrix}=
\frac{1}{2}\begin{pmatrix}1  & i & i & -1\\
			i & 1 & -1 & i\\
			i & -1 & 1 & i\\
			-1 & i & i & 1
			\end{pmatrix},
			\end{equation}
with $t$ and $r$ as defined in Sec.~\ref{sec:intro}. The final state is obtained in a straightforward way by applying the beamsplitter operator to the initial state			
 \begin{equation}
 \ket{\psi}_f=\hat{B}_2\ket{\psi}_i=\frac{i}{\sqrt{2}}\begin{pmatrix}1 \\ 0 \\ 0 \\ 1\end{pmatrix}
 \label{eq:psif}
 \end{equation}
Notice that to within an overall phase, this state is equivalent to
\begin{equation}
\ket{\psi}_f=\frac{1}{\sqrt{2}}\left(\ket{x}_1\ket{x}_2+\ket{y}_1\ket{y}_2\right).
\label{eq:find}
\end{equation}
That is, in the final state both photons end up traveling along the same direction. Should we put photon detectors at each output of the beamsplitter, we would not get any coincidences. Note that Eq.~(\ref{eq:psif}) is a nonseparable (entangled) state of the two photons in the momentum degree of freedom.

If we put detectors at the two outputs of the beamsplitter, the probability of detecting photon 1 in one detector and  photon 2 in the other, and {\em vice versa}, is 
\begin{eqnarray}
P_c&=&P(x_1,y_2)+P(y_1,x_2)\label{eq:pc}\\
P_c&=&|\bra{x}_1\bra{y}_2\ket{\psi}_f|^2+||\bra{y}_1\bra{x}_2\ket{\psi}_f|^2\\
P_c&=&0.
\end{eqnarray}
n our photon counting experiment, when a photon impinges on a detector, the detector outputs a pulse that is sent to an electronic circuit. The circuit is set to record when a photon arrives at each of the two detectors within a certain time window, an event defined to be a coincidence. In our HOM experiment, the time window is about 40 ns and the interference effect results in no coincidences being detected.

Let us also consider the situation when there is no interference, that is, when the photons are distinguishable. For example, suppose we know that one photon arrives before the other because the length of paths of the photons from the crystal to the beamsplitter are distinguishably not the same. Suppose also that the photon arrival time from different paths is still within the experimental coincidence window. What would we measure?

One way to analyze the situation is this: assume photon 1 in  momentum state $\ket{x}_1$ arrives at the beamsplitter distinguishably sooner than photon 2, which is in momentum state $\ket{y}_2$. Then, the system's initial state is $\ket{x}_1\ket{y}_2$ and the final state will be:
\begin{equation}
\ket{\psi}_{f,a}=\hat{B}_2\ket{x}_1\ket{y}_2=\frac{1}{2}\begin{pmatrix}i \\ 1 \\ -1 \\ i\end{pmatrix}
\end{equation}
Thus, from Eq.~(\ref{eq:pc}), the probability of measuring a coincidence experimentally will be $P_c=1/2$. We get the same result when we consider the other possibility, i.e., initial state $\ket{y}_1\ket{x}_2$. To make it symmetric we could say that the first possibility occurs half the time and the second possibility occurs the other half. This still gives us $P_c=1/2$.

Based on the previous discussion contrasting bosons and fermions, we can repeat this analysis for the case of fermions. Then, the initial wavefunction must be antisymmetric [i.e., Eq.~(\ref{eq:ind}) with a minus sign instead of a plus sign] because the total wavefunction must change sign with particle exchange. The application of the beamsplitter operation preserves the symmetry of the initial state, yielding an output state that is antisymmetric (i.e., the same as the initial state), underscoring that identical fermions cannot be in the state of Eq.~(\ref{eq:find}). Thus, the symmetry of the wave function accounts for the way the amplitudes combine in Eq.~(\ref{eq:homint}) (i.e., plus sign for bosons and minus sign for fermions).

\subsection{The Dip}\label{sec:dip}

In parametric down-conversion, photon pairs are emitted with energies $E_1=E$ and $E_2=E_0-E$, where $E_0$ is the energy of the pump (parent) photon. Right before being detected, the photons go through energy filters, which restrict the energy of the photons further. Thus, the apparatus detects photons in a superposition of energy states: 
\begin{equation}
\ket{\psi}=\int dE\;a(E)\ket{E}_1\ket{E_0-E}_2
\label{eq:psimeas}
\end{equation}
where $a(E)$ is a measure of the overall bandwidth of the photons that are being detected. The state of the photons in Eq.~(\ref{eq:psimeas}) is non-separable. That is, the photons are in an entangled state of energy. Their exact energy is unknown to within a range of energies $\Delta E$ determined by $a(E)$ and they constitute a wavepacket. As such, the photons are coherent to within the coherence time $\Delta t\sim h/\Delta E$, where $h$ is Planck's constant. Because filters are specified in terms of the wavelength, we can express the energy bandwidth in terms of the wavelength: $\Delta E=hc\Delta \lambda/\lambda^2$, 
where $c$ is the speed of light in vacuum. Thus, a practical way to express the coherence is in terms of the length of the wavepacket, also known as the coherence length
\begin{equation}
\ell_c=c\Delta t=\frac{\lambda^2}{\Delta\lambda}.
\label{eq:lc}
\end{equation}
If the photons arrive at the beamsplitter within the coherence time, then they can be considered indistinguishable (assuming all other photon properties are identical). An alternative reasoning is to say that the difference in the length of the two paths from the crystal to the beamsplitter is less than the coherence length. What occurs in the intermediate cases? For that we need to go deeper into the  quantum mechanics formalism.

In the previous section we analyzed the two extreme interference situations: the photons are indistinguishable yielding no coincidences, or the photons are distinguishable and coincidences are observed. In the former case, the photons are in a quantum entangled state. In the latter, considering the two possibilities, the photons are in a mixed state. To account for mixed states we need to resort to another quantum-mechanical object for describing the state of the photons---the density matrix.\cite{GalAJP10}

The density matrix for the state of the light in the indistinguishable case is given by the outer product of the vector matrices: 
 \begin{equation}
\hat{ \rho}_{\rm ind}= \ket{\psi}\bra{\psi}=\frac{1}{2}\begin{pmatrix}	0  & 0 & 0 & 0\\
			0 & 1 & 1 & 0\\
			0 & 1 & 1 & 0\\
			0 & 0 & 0 & 0
			\end{pmatrix}.
			\end{equation}
When the photons are in the distinguishable case, the density matrix is the weighted sum of the density matrices for each case considered separately:
\begin{equation}
\hat{\rho}_{\rm xy}=\ket{x}_1\ket{y}_2\bra{y}_2\bra{x}_1
%=\begin{pmatrix}	0  & 0 & 0 & 0\\
%			0 & 1 & 0 & 0\\
%			0 & 0 & 0 & 0\\
%			0 & 0 & 0 & 0
%			\end{pmatrix}.
			\end{equation}
			and 
\begin{equation}
\hat{\rho}_{\rm yx}=\ket{y}_1\ket{x}_2\bra{x}_2\bra{y}_1,
%=
%\begin{pmatrix}	0  & 0 & 0 & 0\\
%			0 & 0 & 0 & 0\\
%			0 & 0 & 1 & 0\\
%			0 & 0 & 0 & 0
%			\end{pmatrix},
			\end{equation}
with the mixed state for the all the cases given by
\begin{equation}
\hat{\rho}_{\rm dis}=\frac{1}{2}\hat{\rho}_{\rm xy}+\frac{1}{2}\hat{\rho}_{\rm yx}=
\frac{1}{2}\begin{pmatrix}	0  & 0 & 0 & 0\\
			0 & 1 & 0 & 0\\
			0 & 0 & 1 & 0\\
			0 & 0 & 0 & 0
			\end{pmatrix},
			\end{equation}

The density matrix after the photons go through the beamsplitter is given by
\begin{equation}
\hat{\rho}_f=\hat{B}\hat{\rho}_i\hat{B}^+.
\end{equation}
Readers can show that this outcome is consistent with $\ket{\psi}_f\bra{\psi}_f$ of Eq.~(\ref{eq:find}).
The coincidence probability will be given by Eq.~(\ref{eq:pc}), which when using the density matrix, is expressed in terms of the trace of the product of the density matrix of the state times the density matrix of the state being measured ($\ket{x}_1\ket{y}_2$ or $\ket{y}_1\ket{x}_2$). This results in
\begin{equation}
P_c={\rm Tr}[\hat{\rho}_{f}\hat{\rho}_{xy}]+{\rm Tr}[\hat{\rho}_{f}\hat{\rho}_{yx}].
\label{eq:tr}
\end{equation}
For the indistinguishable case, we easily find that $\hat{\rho}_f=\hat{\rho}_{\rm ind}$ yields $P_c=0$; for the distinguishable case, $\hat{\rho}_f=\hat{\rho}_{\rm dis}$ yields $P_c=1/2$.

The intermediate case can be expressed by the state%\cite{WernerPRA89}
\begin{equation}
\hat{\rho}_{\rm int}=p\hat{\rho}_{ind}+(1-p)\hat{\rho}_{\rm dis}
\end{equation} 
where $p$ is the probability that the photons are indistinguishable. This state is similar to the form of the Werner state.\cite{WernerPRA89} This matrix describes the situation when the state of the photons is partly indistinguishable and partly distinguishable, with $p$ and $(1-p)$ determining the relative weights.
It is left to the reader to show that the final density matrix after the beamsplitter is
\begin{equation}
\hat{\rho}_{\rm int-f}=\frac{1}{4}\begin{pmatrix}1+p & 0 & 0 & 1+p \\ 
							    0 & 1-p & -1+p & 0 \\
							    0 & -1+p & 1-p & 0\\
							    1+p & 0 & 0 & 1+p\end{pmatrix}
							    \label{eq:rhowf}
\end{equation}

\begin{figure}[h!]
  \centering
\includegraphics[width=3in]{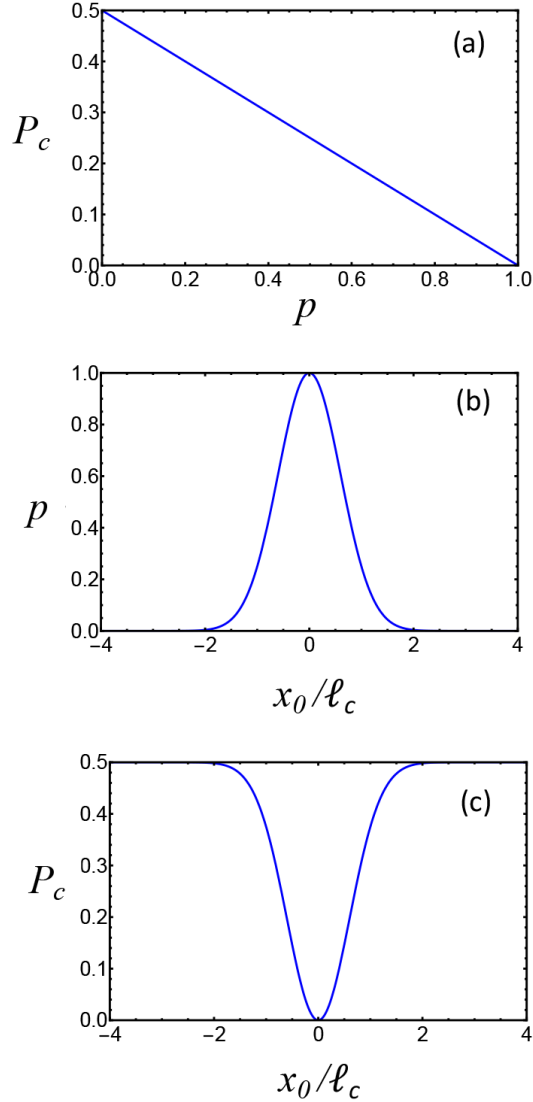}
\caption{(a) Calculated probability of measuring a coincidence  as a function of the Werner probability $p$; (b) Calculated Werner probability using a simple Gaussian model as a function of the the delay in overlap of the photon amplitudes $x_0$ relative to the coherence length $l_c$; (c) Calculated probability of measuring a coincidence  as a function  of $x_0/l_c$. }
\label{fig:HOMp}
\end{figure}

%\begin{equation}
%P={\rm Tr}[\hat{\rho}_{gf}\hat{\rho}_{xy}]+Tr[\hat{\rho}_{gf}\hat{\rho}_{yx}]
%\label{eqdip}
%\end{equation}
 Using $\hat{\rho}_{f}=\hat{\rho}_{\rm int-f}$ as given by Eq.~\ref{eq:rhowf}, Fig.~\ref{fig:HOMp}(a) shows the calculated coincidence probability from Eq.~(\ref{eq:tr}) as a function of $p$.

The landmark experiment by Hong, Ou, and Mandel demonstrated the interference effect by scanning the difference in path, and therefore the overlap of the interference of the two photon amplitudes, exhibiting a famous ``dip'' in the coincidences. We can reproduce the dip analytically using a simple model. If we consider the photon wavepackets as Gaussians with a width at half maximum given by $\ell_c$ but displaced by the path difference $x_0$, then $p$ is proportional to the overlap integral. If we displace the two Gaussians by $x_0$, then
\begin{equation}
p(x_0)=\int_{-\infty}^{+\infty}2\sqrt{\frac{2\ln(2)} {\ell_c\pi}}e^{-4\ln(2)x^2/\ell_c^2}e^{-4\ln(2)(x-x_0)^2/\ell_c^2}dx
\label{eq:px0}
\end{equation}
where we have normalized the overlap such that $p(0)=1$, as shown in Fig.~\ref{fig:HOMp}(b). If we now plot the coincidence probability as a function of $x_0$ we recreate the famous HOM dip, as shown in Fig.~\ref{fig:HOMp}(c).
We note that we present this simple model just to capture the essence of the phenomenon as measured in the laboratory. A more accurate calculation would have to take into account the actual measured bandwidth of the light and other experimental details.

\begin{figure}[h!]
  \centering
\includegraphics[width=4in]{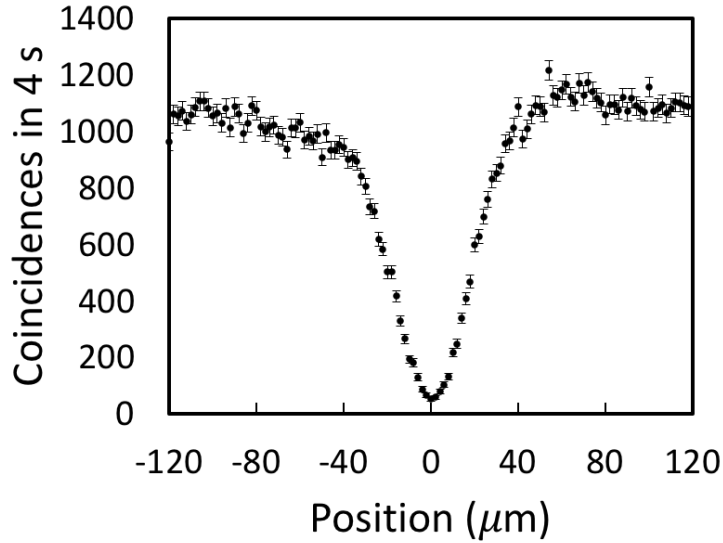}
\caption{Recording of the coincidences as a function of the position of one of the collimators; effectively the difference in the path length of the two photons, from the crystal to the beamsplitter. }
\label{fig:HOMdip}
\end{figure}

After following the alignment procedure outlined in Appendix A and adjusting the waveplates so that photons are input into the fibers with the same polarization, the dip can be found and scanned. In a lab experience lasting only a few hours, the initial alignment is best done for the students beforehand. Perhaps other students can bring the apparatus to this point as part of a several-week laboratory exercise, as we do in the add-on lab of our quantum mechanics course.\cite{URL} Beyond this point students can be asked to ``discover'' the dip, study it in some detail, and investigate the effect of the bandpass filters on the width of the dip.  In Fig.~\ref{fig:HOMdip}, we show the measurement of the dip taken by an undergraduate student (the first author of this paper), who did the experiment as a senior capstone project. This scan of coincidences was taken at 4~s per data point. The horizontal scale is the position of the motorized stage
(a Matlab program to acquire such data is posted in our website, Ref.~\onlinecite{URL}). 
  Data points were taken every 4 stepper-motor steps of the motorized translation stage, which correspond to a motion of the collimator by 5.33 $\mu$m, or a time delay of 18 fs in 
 arrival times of the two photons. Error bars are the standard for Poisson statistics. Accidental coincidences were of the order of 7 counts.   

The quality of the measured dip can be evaluated by the visibility, defined as 
 $v=(N_{\rm max}-N_{\rm min})/(N_{\rm max}+N_{\rm min})$. We actually obtained this value by fitting an inverted Gaussian to the data, giving 
$v=0.93$,  a remarkably good value, which can be attributed largely to the advantage of using the commercial fiber-coupled beamsplitter. The full width of the dip  at half minimum (FWHM) is about 55 $\mu$m, which is of the same order of  the calculated coherence length of 66~$\mu$m.  With a wider filter of (nominal) 30 nm bandwidth, we got 18 $\mu$m, consistent with the calculation ($22\;\mu$m). There is an asymmetry in the shoulders of the dip. We saw different shoulders when we did the experiment with different parameters (filters and position of the collimators). It depends on the shape of $a(E)$ in Eq.~(\ref{eq:psimeas}), which is related to the shape of the transmission curve of each bandpass filter. We also note that the dip appears only in the coincidence counts. The singles counts (i.e., photons detected by each detector separately) are constant throughout the scan. It underscores that it is a two-photon effect.

\subsection{Polarization Distinguishability}\label{sec:pol}
The photons reaching the beamsplitter can be distinguished by other degrees of freedom. That is, the path length difference of the photons arriving at the beamsplitter can be set to zero, while the photons are still distinguishable. One way this can be done is by manipulating their polarization.\cite{Kwiat92} The photons produced by type-I spontaneous parametric down conversion have the same polarization. If we rotate the polarization of one of them by $90^\circ$, then the two photons are distinguishable by polarization and the interference cancellation disappears. If the polarization setting is between $0^\circ$ and $90^\circ$, then the coincidence probability is somewhere in between. 

Because polarization is represented by two-dimensional space, we can incorporate it in the pure-state description of the light. 
The distinction between this and the  situation of Sec.~\ref{sec:dip} is more subtle, involving open and closed quantum systems.\cite{Petruccione,Sales08} Without getting too technical, we proceed by calculating the probability for this situation by fully accounting for polarization in the state of the light. It also provides an additional way to understand two-photon interference.

If we add the polarization degree of freedom for each photon, this doubles the Hilbert space of each photon, and therefore the two-photon system becomes a 16-dimensional Hilbert space. Doing the matrix operations by hand is a bit unwieldy, with 16-element vectors and $16\times16$ operator matrices, but using various software platforms, such as Mathematica or MATLAB, we can do the laborious linear-algebraic steps easily. If we add polarization to the photon's state, then the initial state, where both photons are vertically polarized is given by
\begin{equation}
\ket{\psi}_i=\frac{1}{\sqrt{2}}\left(\ket{x,V}_1\ket{y,V}_2+\ket{y,V}_1\ket{x,V}_2\right),
\end{equation}
where we have now added a label $V$ to the momentum state of each photon in order to specify its polarization. Each product state is of the form
\begin{equation}
\ket{\varphi}=\begin{pmatrix}x \\ y\end{pmatrix}_1\otimes\begin{pmatrix}H \\ V\end{pmatrix}_1\otimes \begin{pmatrix}x \\ y\end{pmatrix}_2\otimes \begin{pmatrix}H \\ V\end{pmatrix}_2
\end{equation}
where $V$ and $H$ specify vertical and horizontal polarization, respectively. In vector form, the initial state would be $\ket{\psi}_i=2^{-1/2}(0\;0\;0\;0\;0\;0\;0\;1\;0\;0\;0\;0\;0\;1\;0\;0)^T$.
Before the photons reach the beamsplitter, they go through half-wave plates. For symmetry, we add one for each momentum input. (Experimentally, two waveplates are needed to keep the optical path of the two photons as close to equal as possible.) The waveplates for the $x$ and $y$ momentum states are oriented by angles $\theta$ and $\phi$ relative to the vertical direction, respectively.  
We can express the operator for a half-wave plate oriented an angle $\theta$ by
\begin{equation}
\hat{W}_\theta=\begin{pmatrix}-\cos2\theta & -\sin2\theta\\ -\sin2\theta & \cos2\theta\end{pmatrix}.
\end{equation}
Because the two waveplates are attached to each momentum state, the operator for the two waveplates acting on the space of the two photons has the form
\begin{equation}
\hat{Z}_{\theta,\phi}=\hat{P}_x\otimes\hat{W}_\theta\otimes\hat{P}_y\otimes\hat{W}_\phi+\hat{P}_y\otimes\hat{W}_\phi\otimes\hat{P}_x\otimes\hat{W}_\theta,
\end{equation}
where $\hat{P}_x=\ket{x}\bra{x}$ and $\hat{P}_y=\ket{y}\bra{y}$ are projection operators for the momentum states. 
It is left as an exercise for the reader to show that the state $\ket{\psi}^\prime$ after the waveplates and before the beamsplitter is
\begin{equation}
\ket{\psi}^\prime=\hat{Z}_{\theta,\phi}\ket{\psi}_i=\frac{1}{\sqrt{2}}\left(\ket{x,2\theta}_1\ket{y,2\phi}_2+\ket{y,2\phi}_1\ket{x,2\theta}_2\right)
\end{equation}
where for simplicity we have labeled the polarization state by the orientation of the polarization relative to the vertical direction.

The beamsplitter acts only on the momentum states, leaving the polarization states unchanged. Its operator is given by
\begin{equation}
\hat{B}_4=\hat{B}\otimes\hat{I}\otimes{B}\otimes\hat{I},
\end{equation}
where $\hat{I}$ is the identity, representing the inaction of the beamsplitter on the polarization degree of freedom. The next steps are mechanical: computing the final state followed by a calculation of the coincidence probability. At this point in our lab program, we ask students not just to perform the calculation, but to devise how to calculate the coincidence probability $P_c$ in the larger space following the prescription of Eq.~(\ref{eq:pc}) and produce the result
\begin{equation}
P_c=\frac{1}{2}\left[1-\cos^2(2\phi-2\theta)\right].
%P_c=\frac{1}{2}\sin^2(2\phi-2\theta).
\label{eq:HOMhwp}
\end{equation}
Thus, the answer depends only on the relative orientations of the two polarizations, which is 0 when they are equal.

\begin{figure}[h!]
  \centering
\includegraphics[width=4in]{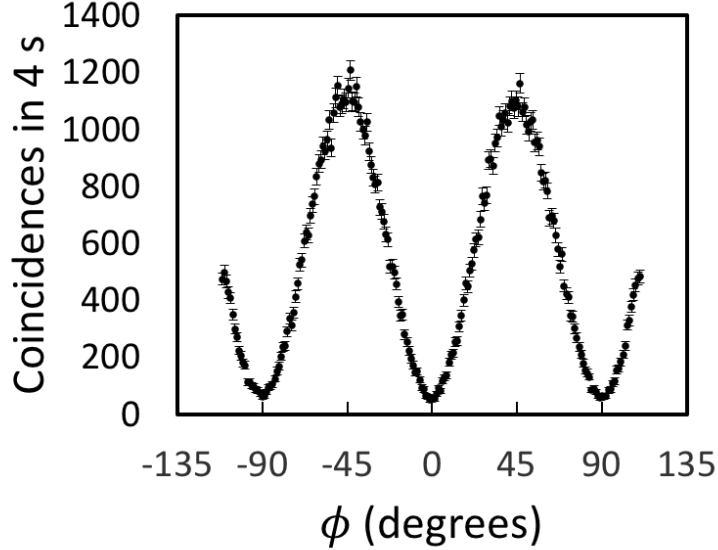}
\caption{Recording of the coincidences as a function of the angular position $\phi$ of one of the half-wave plates, while the other half- wave plate was oriented at $\theta=0^\circ$. Thus, the half-wave plate setting $\phi$ is half the difference in the angular orientation of the polarizations of the two photons. }
\label{fig:HOMhwp}
\end{figure}

Measurements for this section of the experiment follow directly from the setup of the previous ones. Students can also be given the freedom to take the data in whichever form they decide. For example, the data of Fig.~\ref{fig:HOMhwp} shows a scan of the angle $\phi$ of the half-wave plate on the motorized mount. It follows remarkably close to the expectation of Eq.~(\ref{eq:HOMhwp}) for $\theta=0$. At $\phi=\pm 45^\circ$ the two input polarizations are orthogonal, the coincidences are at a maximum, which corresponds to about 1150 counts, the same as the ones corresponding to 1/2 probability in Fig.~\ref{fig:HOMdip}. At $\phi=0, \pm90^\circ$ the polarizations are parallel and we get destructive interference, with about the same  value of counts as at the dip. Because of the simple functional form of the data of Fig.~\ref{fig:HOMhwp}, we fitted Eq.~(\ref{eq:HOMhwp}) with a visibility parameter 
$v$ multiplying the cosine function, to the data (not shown to avoid cluttering the figure), which resulted in an excellent match, with a reduced chi-square of 1.04. The fitted visibility was 
 $v=0.94$, which is quite remarkable for a teaching laboratory experience. 

\section{Conclusions}
In summary, we present a simplified version of the Hong-Ou-Mandel experiment that is suitable for the undergraduate instructional laboratory. The use of a fiber-coupled beamsplitter greatly simplified the alignments. The development of the experiment involved a one-semester senior capstone project. Once we knew how to overcome the challenging parts of the apparatus, disassembly and reassembly proceeded at the same pace as in other single-photon-type of experiments. In Appendix A, we present some of our recommendations for set up and alignment. 
We were quite surprised by the high quality of the data, shown in Figs.~\ref{fig:HOMdip} and \ref{fig:HOMhwp}. Free-space alignment normally produces much lower visibilities, because of difficulties in the alignment. We found that the use of a motorized translation stage greatly streamlined finding the interference dip, and in making detailed scans. The use of waveplates helped in finding the best visibility, which was around 0.93-0.94. Such visibilities make this experiment a strong demonstration of a purely quantum interference effect. The motorized aspect of the scans also allows the experiment to be performed remotely.\cite{GalSPIE}

A discussion of this experiment  leads to fundamental concepts of quantum mechanics, such as  the symmetry of the wavefunction. The lack of coincidence is indeed because the wavefunction of the two photons must be symmetric due to their bosonic nature. Should another part of the wavefunction, such as  polarization\cite{Weinfurter94,Braunstein95} or spatial mode,\cite{Walborn03} be antisymmetric, it would require an antisymmetric momentum wavefunction so that the total wavefunction is symmetric. This situation will lead to a {\em maximum} in the coincidences.\cite{Mattle96,Walborn03} Our classical intuition leads us to associate a physical force (e.g., electromagnetic, gravitational) whenever there is an interaction, but quantum mechanics allows such an interaction between particles to exist just because they are identical. This property also manifests tangibly with (identical) electrons in atoms via the exchange force (see Ref.~\onlinecite{Griffiths} for an illuminating presentation). The same is the case here with photons in a beamsplitter, which is also the basis for using photons in quantum computation.\cite{KLM01} 

The temporal overlap and polarization aspect of the experiments presented above also underscores the requirement of indistinguishability for interference to occur. 
It serves as a basis for discussing another remarkable quantum interference experiment, also known as the ``mind-boggling experiment,'' which exploits the indistinguishability aspects of this interference phenomenon.\cite{Zou91} In that experiment, interference between photons of {\em separate} pairs is seen when the two pairs are indistinguishable from each other, which leads to important consequences for quantum computing purposes, such as entanglement swapping,\cite{Pan98} teleportation,\cite{Bouwmeester97} and the entanglement of multiple qubits. A recent discussion of the more general case of $N$ photons reaching the beam splitter examines other situations not considered in this article.\cite{Masud23}
With quantum mechanics making irreversible inroads into technology, experiments such as this one make students appreciate the inner-workings of quantum mechanics. It constitutes an important step for understanding the technological tools of the future.

\appendix
\section{Experimental Set Up and Alignment}
Setting up the experiment must be done in stages. A first stage involves aligning the crystal and collimators for spontaneous parametric down-conversion. We did this with the simplest of setups. The collimators are attached to multimode fibers. We made marks on the breadboard where the collimators should be placed, 1-m away from the crystal. We used an alignment laser and a homemade plumb bob to mimic the path of the down-converted photons and couple the light into the fibers. Once this was done, we placed 30-nm filters on the collimators, turned the pump laser and the detectors on, and looked for coincidences. After optimization, the singles should be above 10000 counts per second (e.g., for us it was $\sim60000\;{\rm s}^{-1}$ ), and the coincidences between 5\% and 10\%  of the singles counts. 

Once the first stage was completed, we disconnected the multimode fibers and attached the inputs of the fiber-coupled beamsplitter to the collimators. At this point it helped us to couple the light from a laser into a fiber connected to one of the outputs of the beamsplitter. We did this with a commercial low-power handheld fiber-coupled laser, which could also serve as the alignment laser. The purpose of this was to send the light back from the collimators toward the crystal. We used it to adjust the focus of the collimators to match the size of the pump laser on the BBO crystal. This mode-matching step is important in coupling the photons into the  fibers.

After the previous stage was set, we observed singles counts of the order  30,000 counts per second and about 800 coincidences per second. At this point we initiated the search for the dip in coincidences. Doing this by hand was difficult and tedious, mainly because the dip is of the order of 10 $\mu$m wide, so very small steps of the translation stage are needed, and that is difficult to do by hand. Much easier is to do a motorized scan. This finds the dip reliably. We did so both ways, and the motorized way was significantly easier.

Once the dip was found, we switched to narrower bandwidth filters (10 nm), which increased the width of the dip. At this point the dip was not optimally deep. The next and final stage involved adding identical waveplates in front of both collimators. Because they delay the light as it travels through them, we had to find the new location of the dip, displaced by the slight difference in thickness of the two waveplates. Adjustment of the relative orientation of the waveplates at the dip location resulted in the best interference condition. 

\section{Parts List}

Table~\ref{tab:parts} lists the main parts that are needed for this experiment. 
Other parts can be obtained from Ref.~\onlinecite{URL}. 
The central piece for this experiment is the beamsplitter. As mentioned earlier, we used a fiber-coupled beamsplitter, listed in the table. It can also be custom-ordered to a commercial vendor of fiber-optical components, with the specification that it needs equal-length polarization maintaining fibers and use a 50:50 beamsplitter at a wavelength of 810 nm. It can also be done with a fiber beamsplitter,\cite{Qutools} although we have not tried it. 

  \begin{table}[h!]
\centering
\caption{Parts list with vendors, models and rounded prices. Vendor Abbreviations: Newlight Photonics: New. Phot.; Optsigma: Opto.; Power Technology Inc.: Pow. Tech.; Pacific Laser: Pac. Las.. SPADs with educational discount are sold by Alpha.\cite{Alpha} Electronics circuits are based on field programmable gate arrays.\cite{URL,Beck}}
\begin{ruledtabular}
\begin{tabular}{l c p{4.8cm} r p{5cm}}
Name & Number & Vendor \& Model & Price (\$) & Comment \\
\hline	% horizontal line to separate headings from data
Bandpass Filter & 2 & New. Phot. NBF810-30 & 160 & 30 nm, 810-nm center.\\
Bandpass Filter & 2 & New. Phot. NBF810-10 & 160 & 10 nm, 810-nm center.\\
BBO crystal & 1 & New. Phot. NCBBO5300-405(I)-HA3 & 620 & Type-I down-conversion crystal, 5x5x3 mm\\
Beamsplitter & 1 & Qubitekk  
& 3000 & Fiber coupled. \\
Collimator & 2 & Thorlabs  CFC8-B& 300 & Adjustable focus \\
Collimator adapter & 2 & Thorlabs AD15F2 & 30 & For mirror mount.\\
Detectors (SPAD) & 2 & Excelitas SPCM-AQHR & 3000 & Dark counts $\le1000$ cps. \\
Electronics & 1 & Altera DE-115 or Red Dog & 300 & For recording coincidences.\\
Fiber Laser & 1 & Thorlabs HLS635 or OZ Optics FOSS & 700 & 1 mW, hand-held.\\
Filter mount & 2 & Thorlabs SM1L05 & 20 & Mount for filter, 1-inch ID. \\
Iris & 2 & Thorlabs SM1D12 & 60 & Iris mounted on collimator. \\
Mirror mount & 2 & Thorlabs KM100T & 70 & For mounting collimator.\\
Pump laser & 1 & Pow. Tech. GPD405-50 & 510 & 405-nm laser module, 50 mW.\\ 
Rotational mount & 1 & Opto. GTPC-SPH30 & 250 & Manual. \\
Rotational adapter & 1 & Opto. GTPC-ADP25.4-38& 100 & Adapter for 1-inch aperture. \\
Rotational mount & 1 & Pac. Las. RSC-103E & 1600 & Motorized, USB connected. \\
Translation stage & 1 & Thorlabs MT1 & 330 & Manual, with micrometer.\\
Translation stage & 1 & Pac. Las. LST-10L& 1700 & Motorized.\\
Waveplate & 2 & New. Phot. WPA03-H-810 & 330 & half-wave, 810 nm, zero order. \\

\end{tabular}
\end{ruledtabular}
\label{tab:parts}
\end{table}

\begin{acknowledgments}

This work was funded by National Science Foundation grant PHY-2011937. The authors have no conflict of interest.

\end{acknowledgments}

\end{document}